\documentclass{article}
\usepackage{graphicx}  
\usepackage{amsmath}   
\usepackage[compress,numbers,sort]{natbib}
\usepackage{amssymb}   
\usepackage{bm} 
\usepackage{dcolumn}
\usepackage{color}
\usepackage{mathrsfs}
\usepackage{amsfonts}
\usepackage{varioref}
\usepackage{textcomp}
\usepackage[normalem]{ulem}

\usepackage{multirow}
\usepackage{caption}
\usepackage{subcaption}
\RequirePackage[colorlinks,citecolor=blue,urlcolor=magenta,linkcolor=blue]{hyperref}
\allowdisplaybreaks
\labelformat{section}{Section #1} 
\labelformat{subsection}{Section #1} 
\labelformat{subsubsection}{Section #1}
\labelformat{subsubsubsection}{Section #1}
\labelformat{equation}{Eq.~(#1)} 
\labelformat{figure}{Fig.~#1} 
\labelformat{subfigure}{Fig.~\thefigure#1} 
\labelformat{table}{Table~#1} 
\labelformat{appendix}{Appendix #1}

\usepackage{csquotes}
\usepackage{float}
\def\Sech {\rm Sech}
\usepackage[font=footnotesize]{subcaption}
\usepackage{amsmath}
\usepackage{physics}
\usepackage{xspace}
\def\6{\partial}
\def\partialbar{{\mathchar '26\mkern -10mu\partial}}

\title{Effect of dark matter halo environment on GW memory signal }
\author{ Soumya Bhattacharya \footnote{soumya557@gmail.com}$~^{1, 2}$, and  Chiranjeeb Singha\footnote{chiranjeeb.singha@iucaa.in}$~^{2}$\\
$^{1}$\small{Department of Physics, Indian Institute of Technology Bombay, Mumbai 400076, India}\\
$^{2}$\small{Inter-University Centre for Astronomy \& Astrophysics, Post Bag 4, Pune 411 007, India}}
\begin{document}
\maketitle

\begin{abstract}

In this paper, we study the gravitational wave (GW) memory effect for a black hole embedded in a dark matter halo described by a Hernquist-type density profile, both with and without a spike. We first solve the geodesic equations in this spacetime under the influence of a GW pulse to examine how the combined effects of the dark matter halo and the GW pulse modify the geodesic deviation equation and particle trajectories. We then investigate how the memory effect manifests in the waveform in the presence of the dark matter halo. To do that, we analyze the memory contribution at asymptotic null infinity using the Bondi-Sachs formalism and, in particular, the Bondi-Metzner-Sachs (BMS) flux balance laws associated with BMS symmetries. This framework allows us to quantify the GW memory contribution to the waveform, incorporate it into the ringdown waveform templates, and thereby provide a possible avenue for extracting information about the dark matter halo parameters.

\end{abstract}

\section{Introduction}

Numerous observational studies including those examining the flat rotation curves of galaxies \cite{Rubin1970RotationOT, cowie1986virial, Borriello:2000rv, Persic:1995ru}, the dynamics of hot gas in galaxy clusters \cite{Briel:1997hz}, and gravitational lensing effects \cite{SDSS:2005sxd} consistently indicate that approximately 95\% of a galaxy's mass is composed of non-baryonic matter, commonly referred to as dark matter \cite{Clowe:2006eq, Freese:2008cz}. This implies that truly isolated objects are absent in our universe; every compact object, whether a black hole or an exotic compact object (ECO), must coexist with surrounding dark matter, which in turn influences the geometry of spacetime.

In this context, a recent comprehensive study \cite{Cardoso:2021wlq} presented a fully relativistic treatment of a black hole immersed in a galactic matter distribution modeled by a Hernquist-type density profile \cite{hernquist1990analytical}:
\begin{equation}\label{Hernquist-density}
\rho_{\rm h}(r) = \frac{M a}{2 \pi r (r + a)^3}~,
\end{equation}
where $M$ is the total mass of the dark matter halo, and $a$ is a characteristic length scale associated with the dark matter halo distribution. This density profile motivates a corresponding mass function for a galactic black hole, expressed as,
\begin{align}\label{Cardoso-Result-1}
m_{\rm h}(r) = M_{\rm BH} + \frac{M r^{2}}{(r + a)^{2}} \left(1 - \frac{2 M_{\rm BH}}{r} \right)^{2}~,
\end{align}
with $M_{\rm BH}$ denoting the mass of the central black hole. The resulting spacetime geometry closely resembles that of the Einstein cluster model \cite{7bb06a79-8225-31c6-88c3-0c4f8a76b072}. Crucially, this configuration maintains the existence of a black hole horizon, even within the surrounding galactic environment. These findings offer a strong foundation for further exploration into the relativistic modeling of galactic black holes. Here, we also  consider the Hernquist DM spike
profile \cite{Speeney:2024mas, Chakraborty:2024gcr}.
The radial profile of the dark matter (DM) density is described by  
\begin{equation}\label{eq:rho_x}
    \rho_{\rm sp}(x;\lambda) = A(\alpha,\beta,\gamma) \left[\left(1 - \frac{4}{x}\right)^w x^{-q} \left(1 + \frac{x}{x_S^\prime}\right)^{q-4}\right],  
\end{equation}
where $x = 2 r/R_{\rm S}$ is the dimensionless radial coordinate. The normalization constant $A$ depends on three physical scales: the black hole mass $M_{\rm BH}$, the total DM halo mass $M_\mathrm{DM}$, and the physical extent of the halo $r_\mathrm{s}$. The parameter $x_S^\prime$ is defined as  
\begin{equation}\label{eq:xs_prime}
    x_S^\prime = \frac{2r_\mathrm{s}M_\mathrm{DM}}{M_{BH}^2}~.
\end{equation}
It is evident that the density profile $\rho_{\rm sp}$ diverges at $x = 4$, hence, following the prescription in \cite{Chakraborty:2024gcr}, the DM spike is truncated at $x \leq 4$. The analytic expression for the mass profile corresponding to the Schwarzschild–Hernquist spike configuration is given by \cite{Chakraborty:2024gcr},
\begin{equation}
m_{\rm sp}(r) = M_{BH} + \lambda R_S^3 \tilde{A} \, _2F_1\left(w+1, q+w-2; w+2; -\frac{(x - 4)x_S^\prime}{4(x + x_S^\prime)}\right)\Theta(x - 4)~,
\label{mass_SH}
\end{equation}
where $_2F_1$ denotes the Gauss hypergeometric function, and $\Theta$ is the Heaviside step function. Here, $R_{S}=2M$ denotes the Schwarzschild radius. The exact expression for $A(\alpha,\beta,\gamma)$ and $\tilde{A}$ is provided in Eq. (20) of ref. \cite{Chakraborty:2024gcr}. We also adopt the same values of $w$ and $q$ as in ref. \cite{Chakraborty:2024gcr}.

Over the past decade, two major observational breakthroughs have significantly advanced the field of gravitational physics: the direct detection of gravitational waves (GW) from binary black hole and binary neutron star mergers \cite{LIGOScientific:2016aoc, LIGOScientific:2017vwq, KAGRA:2025oiz}, and the imaging of the shadows of supermassive compact objects, such as M87* and SgrA* \cite{EventHorizonTelescope:2019dse, EventHorizonTelescope:2019jan, EventHorizonTelescope:2019pgp, EventHorizonTelescope:2019ths, EventHorizonTelescope:2019uob}. Both phenomena are deeply rooted in the strong-field regime of gravity and, in principle, offer powerful avenues to test the validity of general relativity (GR) and to gain insights into the true nature of compact astrophysical objects.

Here, we aim to explore a relatively uncharted yet highly promising direction for the next generation of GW detectors: the \textit{gravitational wave memory effect}. As detection capabilities improve both through future ground-based observatories and with the anticipated launch of the space-based \textit{LISA} mission, we may finally gain the opportunity to observe this elusive phenomenon \cite{Boersma:2020gxx}. The gravitational memory effect encapsulates both strong-field and non-linear features of general relativity that have yet to be directly observed. It manifests as a permanent displacement between test particles after a GW passes through spacetime \cite{Braginsky:1985vlg, Favata:2010zu}. This subtle, DC-like shift in the GW amplitude has so far evaded detection by observatories such as LIGO \cite{Hubner:2021amk}, primarily due to its weak signature. To enhance the chances of detection, methods such as stacking multiple GW signals from LIGO-Virgo have been proposed~\cite{Lasky:2016knh}.

Initially studied in contexts like hyperbolic scattering \cite{Zeldovich:1974gvh} and gravitational bremsstrahlung~\cite{Kovacs:1978eu}, the memory effect has since been investigated in a wide array of settings \cite{Christodoulou:1991cr}. These include extensions to classical electrodynamics \cite{Bieri:2013hqa, Winicour:2014ska}, Yang-Mills theories \cite{Pate:2017vwa, Jokela:2019apz}, and scenarios involving extra spatial dimensions 
\cite{Hollands:2016oma, Satishchandran:2017pek, Ferko:2021bym}. 
It has also served as a diagnostic tool for distinguishing general relativity (GR) from alternative theories of gravity, such as scalar-tensor models \cite{Du:2016hww, Hou:2020tnd, Hou:2020xme, Tahura:2020vsa} and Chern-Simons gravity \cite{Hou:2021oxe}. For various black hole mimicker spacetimes in GR and beyond GR \cite{Bhattacharya:2022, bhattacharya:2023, bhattacharya:2025}, for neutrino self-interaction of supernova \cite{Bhattacharya_2024}, and various other theoretical aspects \cite{Zhang:2017, Zhang:2017soft, Horvathy_2017}. Furthermore, recent studies have extended the notion of memory to symmetries near black hole horizons \cite{Donnay:2018ckb, Bhattacharjee:2019jaf, Rahman:2019bmk}.

In this work, we investigate the GW memory effect in a more astrophysically relevant setting: a black hole situated at the center of a galaxy, embedded in a surrounding dark matter halo described by a Hernquist-type density profile, both with and without a spike. This configuration introduces a rich dynamical interplay between the central black hole, the background matter distribution, and passing GW pulses. Recent works have demonstrated that astrophysical environments can imprint transient signatures, such as amplitude modulations, in both the waveform tails and the (linear) memory \cite{Alnasheet:2025mtr}. In this study, we pursue a different direction. Specifically, we first solve the geodesic equations in a spacetime influenced simultaneously by the central black hole and an incident GW burst \cite{Braginsky:1985vlg, Zhang:2017geq}. This framework enables us to examine how the geodesic deviation equation and particle trajectories differ from the idealized isolated black hole case, thereby uncovering the effects of the dark matter halo on GW memory observables.

Then, to understand the asymptotic properties of this spacetime and rigorously define the memory effect, we adopt the Bondi-Sachs formalism and study the behavior of the metric and associated fields at null infinity \cite{Madler:2016xju}. Leveraging the BMS symmetry group, we use flux-balance laws to compute the memory effect on the total GW signal. This framework provides a systematic way to incorporate memory signatures into he ringdown waveform templates, thereby improving the prospects for detecting memory effects and offering a new probe of galactic parameters, such as the mass and characteristic length scale of the surrounding dark matter halo.  In constructing memory corrected total waveform templates, we restrict our attention to the dark matter profile without a spike, as the asymptotic properties of spacetimes with and without a spike are identical.

Our study thus aims to bridge theoretical insights from gravitational memory with observational prospects in the context of galactic black holes, contributing to the ongoing effort to characterize the astrophysical and cosmological environment through precision GW measurements.

This paper is organized as follows. In \ref{geodesic}, we solve the geodesic equations in the considered spacetime under the influence of a GW pulse to investigate how the combined effects of the dark matter halo and the GW pulse modify the geodesic deviation equation. In \ref{particle orbit}, we study the motion of a test particle in this spacetime subject to a GW pulse, focusing on the modifications to its trajectory due to the halo and GW pulse. In \ref{bondi-sachs}, we analyze how the memory effect manifests in the ringdown waveform in the presence of the halo. For this purpose, we evaluate the memory contribution at asymptotic null infinity using the Bondi–Sachs formalism, with particular emphasis on the BMS flux balance laws associated with BMS symmetries. Finally, in \ref{conclusion}, we summarize our findings and outline possible directions for future work.

\textbf{\textit{Notations and Conventions}}:  
Throughout this paper, we adopt the mostly-plus signature convention, so that the Minkowski metric in $1+3$ dimensions is expressed in Cartesian coordinates as $\mathrm{diag}(-1,+1,+1,+1)$. We also employ geometrized units, setting $G = c = 1$ throughout.

\section{Memory effect through geodesic analysis}\label{geodesic}
\noindent In this section, we analyze the memory effect in the context of geodesic deviation between neighboring geodesics induced by a passing GW. The separation between geodesics serves as a measure of the displacement memory effect. Furthermore, if the geodesics do not maintain a constant separation after the GW pulse has passed, a velocity memory effect can also be attributed to them though we are not analysing the velocity memory in this work.
We will use the Bondi-Sachs coordinates to perform the memory analysis. Let us introduce the Bondi-Sachs coordinates and write down the metric line element in these coordinates. We employ the coordinate transformation $u = t - r_*$, where $r_*$ denotes the tortoise coordinate, defined through $(dr_*/dr) = (1/\sqrt{-g_{tt} g^{rr}})$.
The line element of the galactic black hole in these coordinates takes the following form for both a spike and without a spike DM profile,
 \begin{equation}
     ds^2 = -f(r) ~du^2 - 2 \sqrt{\frac{f(r)}{g(r)}}~du dr + r^2 ~d\Omega_2^2 ~,\label{BSMi}
 \end{equation}
 where,
 \begin{eqnarray}
     f(r)=f_{\rm h}(r) = \Big(1-\frac{2M_{\rm BH}}{r}\Big) e^\gamma, ~~~~~g(r)=g_{\rm h}(r) = 1 - \frac{2 m_{\rm h}(r)}{r}, \\
     \gamma = -\pi \sqrt{\frac{M}{\xi}} + 2 \sqrt{\frac{M}{\xi}} \arctan{\frac{r + a -M}{\sqrt{M \xi}}}, \\
     \xi = 2 a - M + 4 M_{\rm BH}, \\
     m_{\rm h}(r)= M_{\rm BH}+\frac{M r^{2}}{(a+r)^2}\bigg(1- \frac{2 M_{\rm BH}}{r}\bigg)^{2}\, \, ,
 \end{eqnarray}
corresponding to the Hernquist-type density profile \cite{Cardoso:2021wlq}. If we consider the Hernquist-type dark matter spike profile, the associated mass function $m_{\rm sp}(r)$ is given in \ref{mass_SH}. Utilizing this form, the metric function $f_{\rm sp}(r)$ can be obtained by solving the following differential equation \cite{Chakraborty:2024gcr},
\begin{equation}
\frac{r f'_{\rm sp}(r)}{f_{\rm sp}(r)}=\frac{2 m_{\rm sp}(r)}{r-2m_{\rm sp}(r)}\,.
\label{Diff_eq_gtt}
\end{equation}
 
\noindent Then $f(r)=f_{\rm sp}(r)$ and $g(r)=1-2m_{\rm sp}(r)/r$ corresponding to Hernquist-type dark matter spike profile. The line element with the TT-gauge perturbation for both a spike and without a spike DM profile will look like
\begin{eqnarray}
     ds^2 = -f(r) ~du^2 - 2 \sqrt{\frac{f(r)}{g(r)}}~du dr + r^2( 1 - H(u))~d\theta^2  + r^2(1 + H(u)) \sin^2 \theta~d\phi^{2}~. ~~~~\label{BSMip}
 \end{eqnarray}
 The corresponding geodesic equations in the equatorial plane ($\theta = \pi/2$) will take the following form
 \begin{eqnarray}
     \Ddot{u} - \frac{f'}{2} \sqrt{\frac{g}{f}}\Dot{u}^2 + r (1+ H(u)) \sqrt{\frac{g}{f}} \Dot{\phi}^2 = 0~,\label{ueq}
 \end{eqnarray}
 \begin{eqnarray}
     \Ddot{r} + \sqrt{\frac{g}{f}} \Big(\6_r \bigg(\sqrt{\frac{f}{g}}\bigg)  \Big) \Dot{r}^2 +\sqrt{\frac{g}{f}} f' \Dot{r}\Dot{u} +\frac{g f'}{2} \Dot{u}^2 + \frac{r^2 \sqrt{g} H'(u)-2r(1+ H(u))g\sqrt{f} }{2 \sqrt{f}} \Dot{\phi}^2 = 0~,\label{req}
 \end{eqnarray}
 \begin{eqnarray}
     \Ddot{\phi} + \frac{2}{r} \Dot{r} \Dot{\phi} +\frac{H'(u)}{1 + H(u)} \Dot{u}\Dot{\phi} =0~.\label{pheq}
 \end{eqnarray}
 Here, the `overdot' denotes differentiation with respect to the proper time $ \tau$ associated with the geodesics, while the ‘prime’ indicates differentiation with respect to the argument of the respective function. For example,  $ H'(u) \equiv \frac{dH}{du} $ and $ f'(r) \equiv \frac{df}{dr} $.

\noindent
We have numerically solved the three geodesic equations presented in \ref{ueq}-\ref{pheq} using the symbolic computation software \textit{Mathematica}, and have extensively analyzed the resulting solutions. Specifically, we begin by considering two nearby geodesics in the galactic black hole spacetime and investigate the evolution of their coordinate separation.

\begin{figure}[!htbp]
\includegraphics[scale=0.4]{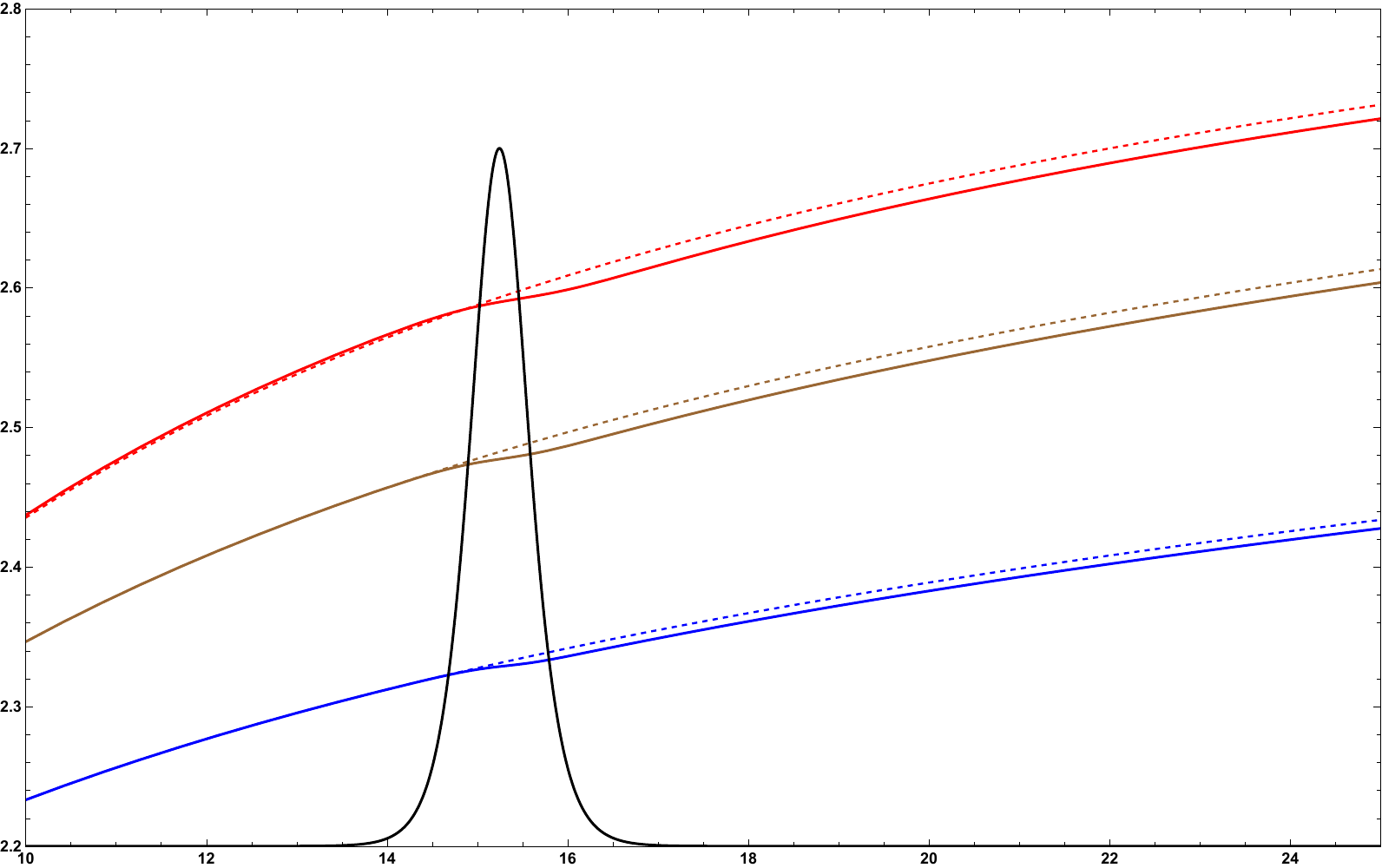}
\put(-320,150){$\bf {\phi}$}
\put(-10,-10){$\bf {\tau}$}
\caption{\raggedright Plot showing the evolution of the geodesic $\phi$ in the presence (solid lines) and absence (dashed lines) of a GW pulse, for a dark matter halo with $M/a = 0.1$ (red), $M/a = 0.3$ (brown), and $M/a = 0.5$ (blue).}
\label{memory_phi}
\end{figure}

\begin{figure}[!htbp]
\includegraphics[scale=0.5]{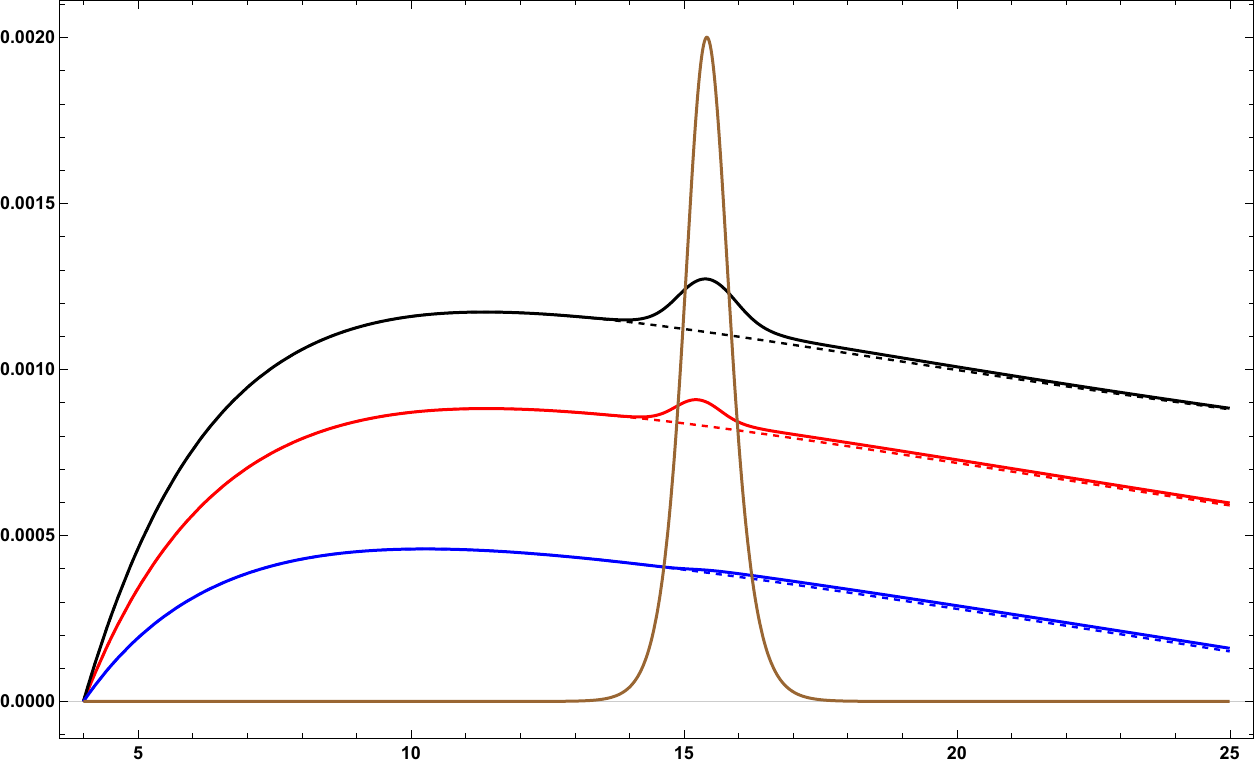}
\put(-320,150){$\bf {\Delta \phi}$}
\put(-10,-10){$\bf {\tau}$}
\caption{\raggedright Plot showing the evolution of the separation $\Delta\phi$ (as defined in \ref{Delta_def}) in the presence (solid lines) and absence (dashed lines) of a GW pulse, for a dark matter halo with $M/a = 0.1$ (black), $M/a = 0.3$ (red), and $M/a = 0.5$ (blue).}
\label{memory_p}
\end{figure}

\begin{figure}[!htbp]
\includegraphics[scale=0.5]{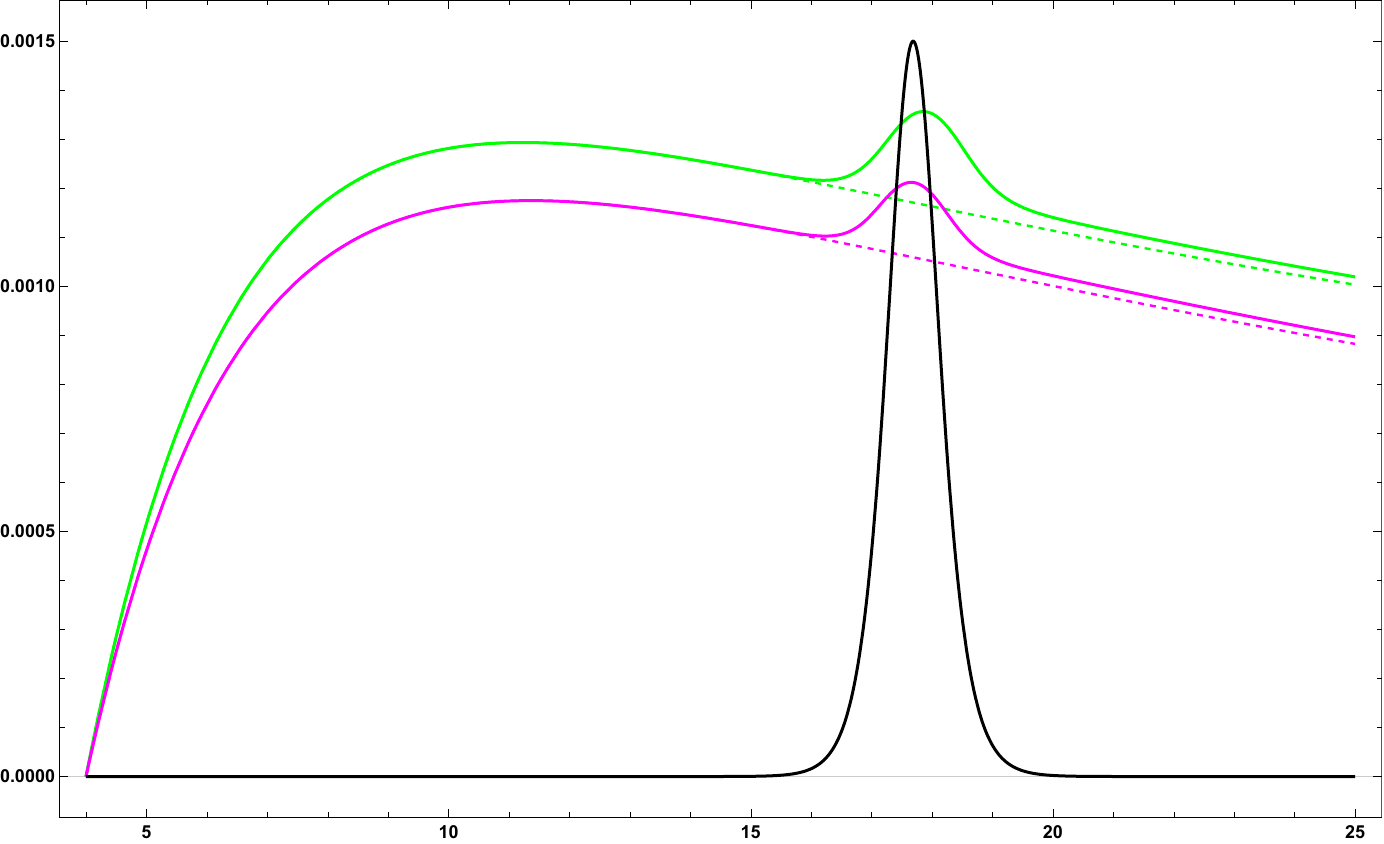}
\put(-350,160){$\bf {\Delta \phi}$}
\put(-10,-10){$\bf {\tau}$}
\caption{\raggedright Plot of the evolution of separation $\Delta \phi$ (as defined in  \ref{Delta_def}) in presence of DM with spike (Green) and DM without spike (Magenta) for $M/a=0.1$, where the solid lines represent the evolution in presence of GW pulse and the dashed lines represent the evolution in absence of GW pulse.}
\label{memory_spike_phi}
\end{figure}

To explore the memory effect, we analyze the evolution of geodesic separation along the $\phi$ coordinate. Schematically, this can be expressed as:

\begin{eqnarray}
   \Delta \phi \equiv \phi({\rm Geodesic~II}) - \phi({\rm Geodesic~I}).~ \label{Delta_def}
\end{eqnarray}
Here, we study the evolution of the quantities $\Delta \phi$ both in the presence and absence of a GW pulse. In the \ref{memory_p} and \ref{memory_spike_phi}, we have plotted the evolution of $\Delta \phi$ in both the presence and absence of a gravitational wave (GW) pulse, considering dark matter halos with and without a spike. In \ref{memory_p}, we have shown the evolution of $\Delta \phi$ in the presence (solid curve) and in the absence (dashed curve) of a GW pulse for different values of $M/a$. In \ref{memory_spike_phi}, we have done a similar analysis as before, but here we consider the dark matter spike profile and compare this scenario with the dark matter halo profile without a spike.  We observe that the behaviour of $\Delta \phi$ is significantly influenced by the presence of the dark matter halo, and is sensitive to its density profile. So we can safely conclude that the memory effect depends on the dark matter halo properties and can act as a pointer to discriminate between various dark matter profiles. Also, one can clearly notice that as $M/a$ increases, the amount of memory decreases, as has been shown in the evolution of the $\phi$ geodesic in the presence and absence of a GW pulse as shown in \ref{memory_phi} and in the evolution of the separation of two nearby $\phi$-geodesics as depicted in \ref{memory_p}.  In the next section, we investigate the memory effect through particle orbits and examine how the dark matter halo influences their trajectories. 
\label{memory_spike_dphi}
\section{Memory effect through particle orbit analysis} \label{particle orbit}

\begin{figure}[!htbp]
    \centering
    \begin{subfigure}[b]{0.48\textwidth}
        \centering
        \includegraphics[scale=0.55]{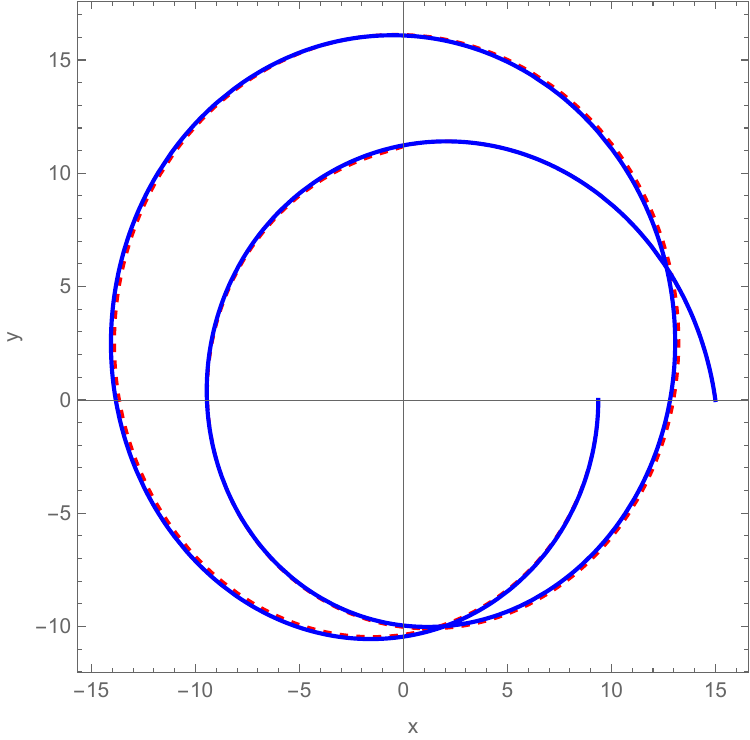}
        \caption{\raggedright Trajectory before and after GW pulse. Blue (thick) is before, red (dashed) is after.}
        \label{Trajectory_1}
    \end{subfigure}
    \hfill
    \begin{subfigure}[b]{0.48\textwidth}
        \centering
        \includegraphics[scale=0.4]{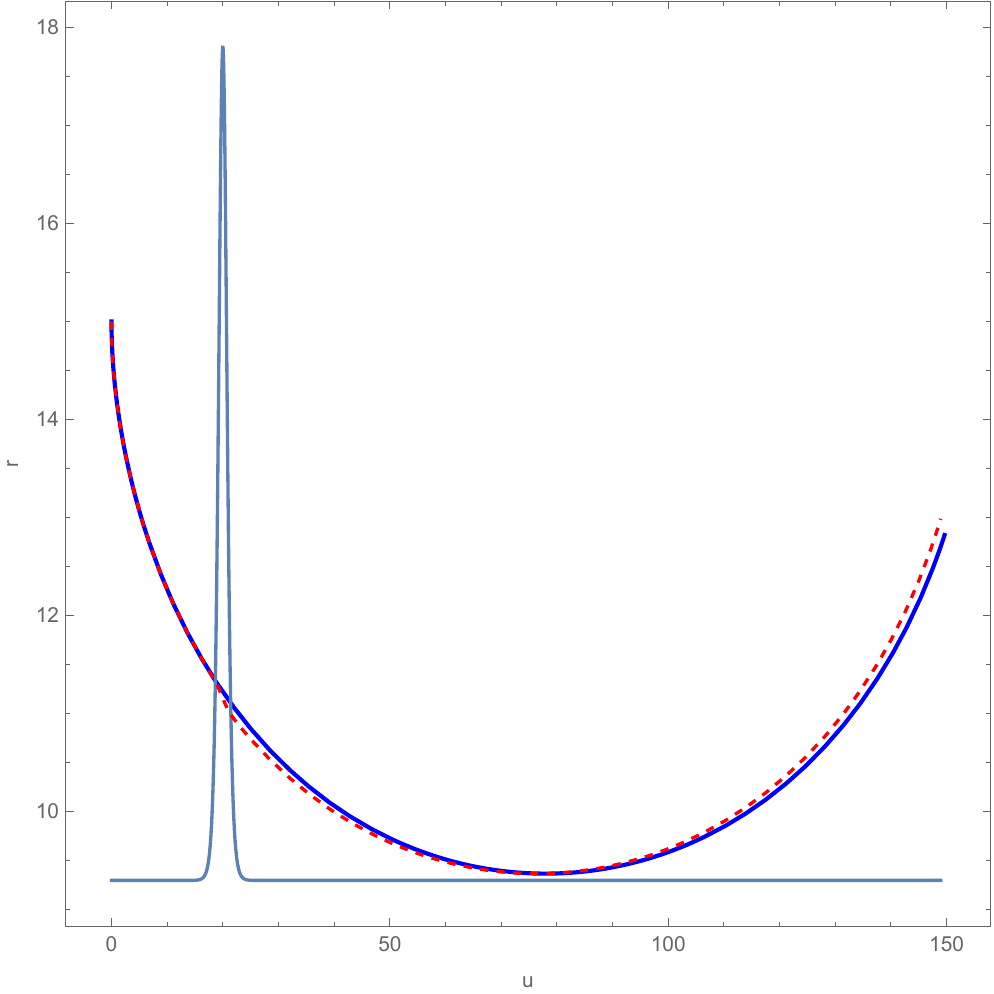}
        \caption{\raggedright Trajectory before and after GW pulse. Blue (thick) is before, red (dashed) is after. The black line is the GW pulse.}
        \label{Trajectory_2}
    \end{subfigure}
    \caption{Comparison of trajectories before and after the GW pulse.}
    \label{fig:Trajectories}
\end{figure}

In this section, we analyze the motion of a test particle in the spacetime geometry surrounding a black hole located at the center of a galaxy. We aim to investigate how the presence of a galactic environment, specifically the influence of the dark matter halo and the passage of gravitational waves (GWs), modifies the particle's trajectory compared to the case of an isolated black hole. We introduce a gravitational wave pulse and observe the resulting deviation in the particle's path. The trajectory alteration persists even after the GW has passed, indicating that the particle retains a memory of the wave, a manifestation of the gravitational wave memory effect. Furthermore, we explore the impact of a surrounding dark matter halo on the dynamics of the particle. This extended matter distribution introduces additional gravitational potential, further modifying the particle’s orbit. We begin by considering the trajectory of a test particle in the Schwarzschild geometry, representing an isolated, spherically symmetric black hole. The timelike geodesic equations in this spacetime can then be written as
\begin{eqnarray}
\dot{u}&=&\frac{E-\dot{r}}{f}~,\nonumber\\
\dot{\phi}&=&\frac{L}{r^{2}}~,\nonumber\\
\dot{r}^{2}&=&E^{2}-f(1+\frac{L^2}{r^2})~.
\end{eqnarray}
Here $E$ is the energy of the particle, $L$ is the angular momentum of the particle, and $f=(1-2M/r)$, ‘overdot’ denotes the derivative
with respect to the affine parameter $\tau$. If we define $U=1/r$, then the equation governing the particle trajectory can be written as,
\begin{equation}
\frac{d^2 U}{d \phi^{2}}+U=\frac{M}{L^{2}}+3 M U^{2}~.
\end{equation}
Now we consider a GW pulse as
\begin{equation}
H(u)= A ~\Sech^2 (u-u_0)~.
\end{equation}
We next evaluate the particle trajectory after the GW pulse by incorporating the corresponding modifications to the timelike geodesic. The trajectories before and after the GW pulse are shown in \ref{Trajectory_1} and \ref{Trajectory_2}, respectively. As illustrated in \ref{Trajectory_2}, the GW pulse leads to a clear deviation in the particle’s trajectory.

\begin{figure}[!htbp]
    \centering

    \begin{subfigure}[b]{0.48\textwidth}
        \centering
        \includegraphics[scale=0.5]{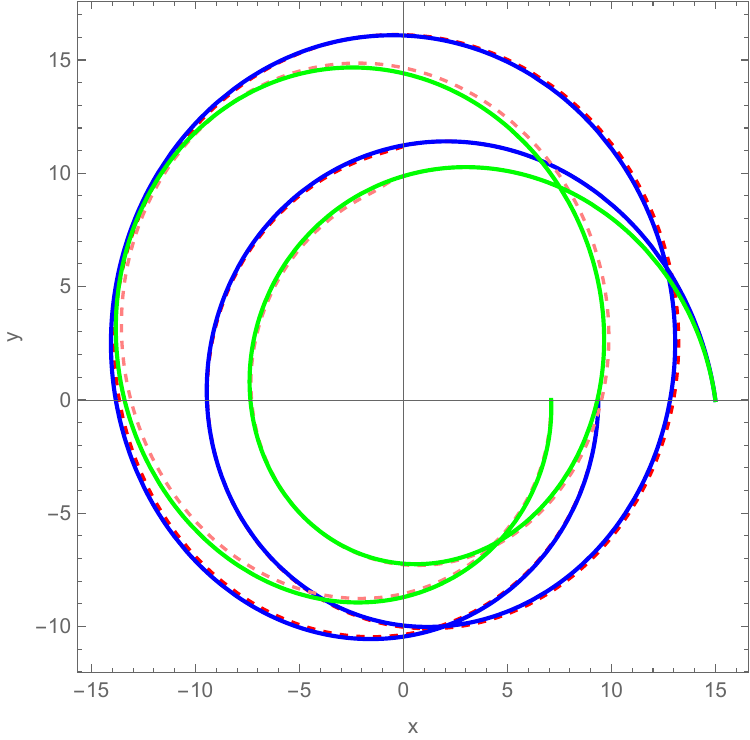}
        \caption{\raggedright The blue, thick line is before the GW pulse, and the red dashed one is after in the absence of a dark matter halo. Similarly, the green, thick one is before the GW pulse, and the pink dashed one is after the GW pulse in the presence of a dark matter halo. Here we have considered $M/a=0.1$.}
        \label{Trajectory_3}
    \end{subfigure}
    \hfill
    \begin{subfigure}[b]{0.5\textwidth}
        \centering
        \includegraphics[scale=0.4]{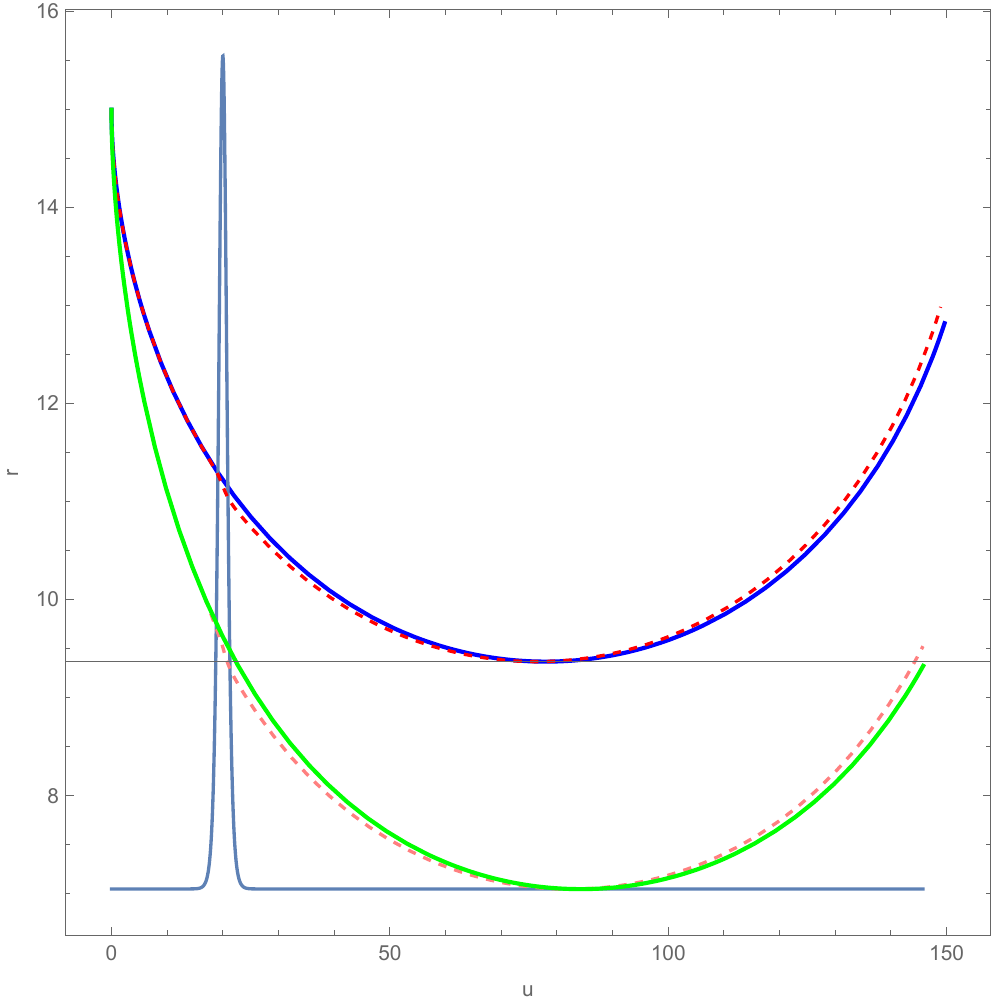}
        \caption{\raggedright Same as (a), with the black line indicating the GW pulse.}
        \label{Trajectory_4}
    \end{subfigure}

    \caption{Plots of the trajectory before and after the GW pulse in the absence and presence of a dark matter halo.}
    \label{fig:Trajectory_combined}
\end{figure}

\begin{figure}[H]
\centering
\includegraphics[width=0.5\textwidth]{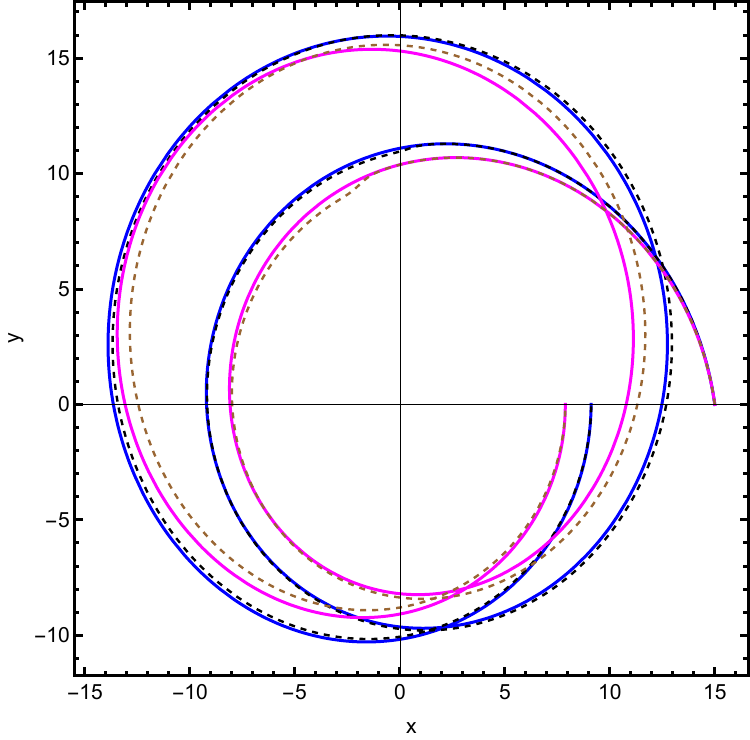}
\caption{\raggedright Plot of the trajectory in the presence of DM spike (black-blue combination) and in the absence of DM spike (magenta-brown combination) for $M/a = 0.1$, where in each case the solid lines represent the trajectory without any GW pulse, and the dashed lines represent the same with a GW pulse. These trajectories clearly show the presence of a memory effect, and it also depends on the DM profile.}
\label{Trajectory_4_1}
\end{figure}

We then examine how the presence of the dark matter halo modifies the trajectory. Using the metric \ref{BSMi}, we derive the corresponding timelike geodesic equations, though their explicit forms are omitted here due to their length. The resulting trajectories in the presence and absence of the dark matter halo are shown in \ref{Trajectory_3} and \ref{Trajectory_4}. We observe that the presence of the dark matter halo modifies the particle's trajectory. We also consider the DM spike profile. In \ref{Trajectory_4_1}, we have plotted the trajectory in the presence and absence of a DM spike. It is evident that the dark matter halo profile distinctly modifies the trajectory.  Though the memory effect is better understood when we consider the separation between two trajectories, here, at least one can see from these trajectory plots that the memory effect is quite susceptible to the presence of dark matter halo as well as to different dark matter profiles. Exact quantification may not be done from here, and hence, we discuss the memory effect in the next section through the paradigm of waveform analysis.

\section{BONDI-SACHS FORMALISM AND MEMORY
 EFFECT FROM NULL INFINITY} \label{bondi-sachs}
 \noindent  We now focus our attention on investigating the
 memory effect at null infinity. As in the previous sections, here also we
 assume that the spacetime of interest corresponds to a
 background spacetime containing a black hole binary immersed in a dark matter halo. A GW pulse passes through it, ultimately reaching the future null infinity, and as
 a consequence, modifying the Bondi mass aspect. 
 So far, we have not imposed any sort of asymptotic flatness condition. Any geometry can be described in a similar form. Imposing asymptotic flatness at large $r$ with fixed $(u, x^A)$ leads to some fall-off conditions of the metric components which give rise to the so-called Bondi gauge conditions \cite{Strominger:2017zoo}. 
 \noindent Let us now consider our case and see how the metric components behave at large $r$. 
 \begin{eqnarray}
     g_{uu} = -f = -1 + \frac{2 (M_{BH} + M)}{r} - \frac{4 M_{BH} M + 2 a M}{r^2}   + \mathcal{O}\Big(\frac{1}{r^3} \Big)~. ~~~ \label{f}
 \end{eqnarray}
 \begin{eqnarray}
     g_{ur} = -1 + \frac{(M+ M_{BH})^2}{r^2} + \mathcal{O}\Big(\frac{1}{r^3} \Big)~. \label{g}
 \end{eqnarray}
We can clearly see from \ref{f} and \ref{g} that these metric components fall off according to the Bondi gauge.
 So, the large $r$ structure of the metric line element will be of the following form: 
 \begin{eqnarray}
     ds^2 = -du^2 -2 ~du dr + r^2 \gamma_{AB}~dx^A dx^B ~~~ {\rm (Minkowski)} \nonumber~~ \\ + \frac{2 (M_{BH} + M)}{r} ~du^2 + \mathcal{O}(r^{-2}) ~~(1/r ~{\rm correction})~.\nonumber\\\label{ABSM}
 \end{eqnarray}
 Here $\gamma_{AB}$ is the metric on the unit two-sphere.  As evident, the Bondi mass aspect is
 simply given by $M_B = (M_{BH} + M)$  and is a constant for the
 background spacetime. This provides the behavior of the
 background static geometry at future null infinity. \\
 \noindent  The above analysis is about the background spacetime,
 which is definitely non-radiative. This is because the metric
 given in \ref{ABSM}  has a constant and non-dynamical Bondi
 mass. This is because there is no loss of news in the absence of
 any dynamics that fit for a non-radiative geometry. Thus, the memory effect requires a propagating GW pulse on top of this background geometry, which leads to a finite radiative term.
 The perturbed line element looks like
 \begin{eqnarray}
ds^2 &=& -du^2 - 2 du dr+\bigg(\frac{2(M_{\rm BH} + M)}{r}  
+ \frac{2M_{\rm B}(u, x^A)}{r} \bigg)du^2 -\bigg(\frac{(M+ M_{BH})^2}{r^2} \nonumber\\&+&\frac{1}{16r^2}C_{AB}(u,x^A)C^{AB}(u, x^A)\bigg) du dr  
+D^B C_{AB}(u,x^A) du\, dx^A 
  +r^2 \bigg(\gamma_{AB} \nonumber \\&+& \frac{C_{AB}(u,x^A)}{r} + \mathcal{O}(r^{-1})  \bigg) dx^A dx^B + \cdots~ \,\, .
\end{eqnarray}
The $1/r^2$ part of the $uu$ component of the Einstein field equations will give rise to the evolution equation of the Bondi mass aspect $m_B(u,x^A)$ as
\begin{eqnarray}
    \6_u m_B(u, x^A) &=& \frac{1}{4} D^A D^B N_{AB}(u, x^A)- \frac{1}{8} N^{AB}(u, x^A) N_{AB}(u, x^A)
    \label{BMe}~.
\end{eqnarray}
The quantity $N_{AB}(u, x^A)$ is called the Bondi news tensor which is defined as $N_{AB} = \6_u C_{AB}$ and $C_{AB}(u, x^A)$ is called the Bondi-shear. 
\\
\noindent We will now focus on the construction of the GW memory waveform based on the above formalism. We know that gravitational waveforms play a crucial role in comparing observed signals to theoretical predictions. The derivation of analytical waveforms directly from GR often poses formidable challenges. However, in a recent work \cite{Ashtekar_2020}, a new avenue opens up which is founded on the balance laws derived directly from complete, non-linear GR. Hence, these balance laws and their utility in
evaluating waveform models are key focal points in our present work. In recent times, various avenues have been studied using this formalism in \cite{Mitman_2020, Mitman_2021, deppe2025}. Particularly in \cite{Mitman_2021}, it is described nicely how to correct the strain waveforms in the Simulating eXtreme Spacetimes (SXS) Collaboration’s catalogue to include the missing displacement memory effect using the BMS flux balance law. So, without much further ado, let us discuss briefly the BMS flux balance law and how to include the memory effect using it in the waveform. \\
The BMS charges and BMS fluxes are related through the mass loss equation given in \ref{BMe}. We can rewrite this equation in terms of the GW strain as follows
\begin{equation}
    \dot{m_B} = -\frac{1}{4} \dot{h}\dot{\Bar{h}} + \frac{1}{4} {\rm Re}(\partialbar ^2 \dot{h}) \, . \label{mbdot}
\end{equation}
Since we would like to compute memory in terms of the strain $h$, we now rewrite the Bondi mass loss equation in terms of strain $h$ by using the following relation between strain and Bondi-shear
\begin{equation}
    h = q^A q^B C_{AB}~,
\end{equation}
where $q^A, ~q^B$ are the dyads with respect to the two-sphere metric $\gamma_{AB}$ and $\partialbar$ is the differential spin-weight operator. As described in \cite{Ashtekar_2020, Mitman_2021}, we can rephrase \ref{mbdot} in the following form
\begin{equation}
    \partialbar^2 h = \int_{-\infty}^u |\dot{h}|^2 du - 4 \Big( \Psi_2 + \frac{1}{4} \dot{h} \Bar{h}   \Big) - 4 M_{\rm ADM} \, , \label{delh}
\end{equation}
where $h \to 0$ as $u \to -\infty$ and $M_{\rm ADM}$ is the ADM mass of the system. 
From \ref{delh} we get the expression of the memory strain as 
\begin{equation}\label{memory}
    h^{\rm memory}= \frac{1}{2} \Bar{\partialbar}^2 \mathcal{D}^{-1} \Big[\frac{1}{4} \int_{-\infty}^u |\dot{h}|^2 du - \Big( \Psi_2 + \frac{1}{4} \dot{h} \Bar{h}   \Big)\Big] \, ,
\end{equation}
here $\mathcal{D} = \frac{1}{8} D^2(D^2 +2)$ with $D = \Bar{\partialbar} \partialbar$, and $\Psi_2$ is the Weyl scalar which represents some radiative degrees of freedom and is related to the integrated Bondi mass as \cite{Mitman_2020, Moxon_2020}
\begin{equation}
    m_B = -{\rm Re}\Big(\Psi_2 + \frac{1}{4} \dot{h} \Bar{h}  \Big) \, .
\end{equation}
In our case, $\Psi_2=0$.  Now the memory corrected total waveform will be 
\begin{equation}
    h^{\rm total} = h + h^{\rm memory} \, . \label{tatal_wavefom}
\end{equation}

\begin{figure*}[htbp]
    \centering
   \begin{subfigure}[b]{0.30\textwidth}
    \centering
    \includegraphics[width=\linewidth]{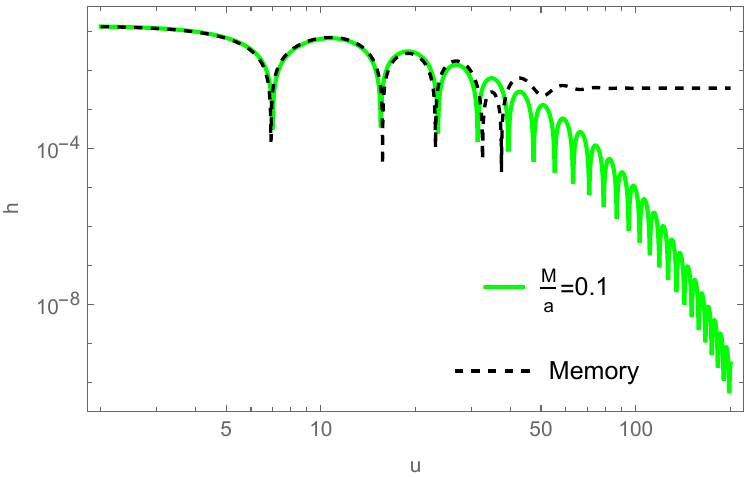}
    \caption{$M/a = 0.1$}
    \label{fig:Memory01}
\end{subfigure}
\hspace{0.02\textwidth}
\begin{subfigure}[b]{0.30\textwidth}
    \centering
    \includegraphics[width=\linewidth]{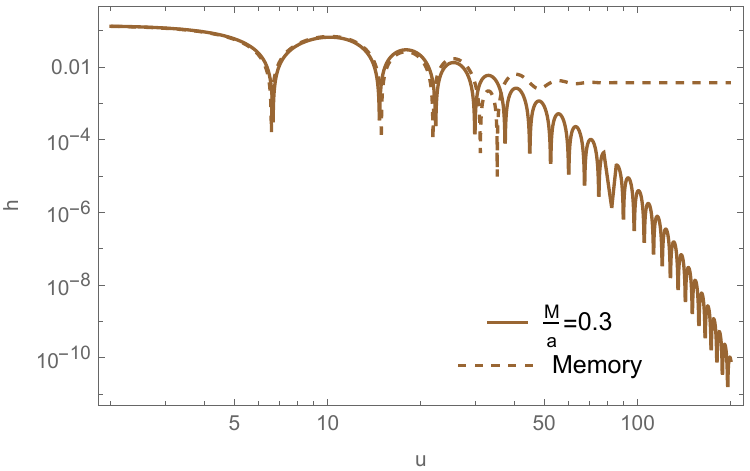}
    \caption{$M/a = 0.3$}
    \label{fig:Memory03}
\end{subfigure}
\hspace{0.02\textwidth}
\begin{subfigure}[b]{0.30\textwidth}
    \centering
    \includegraphics[width=\linewidth]{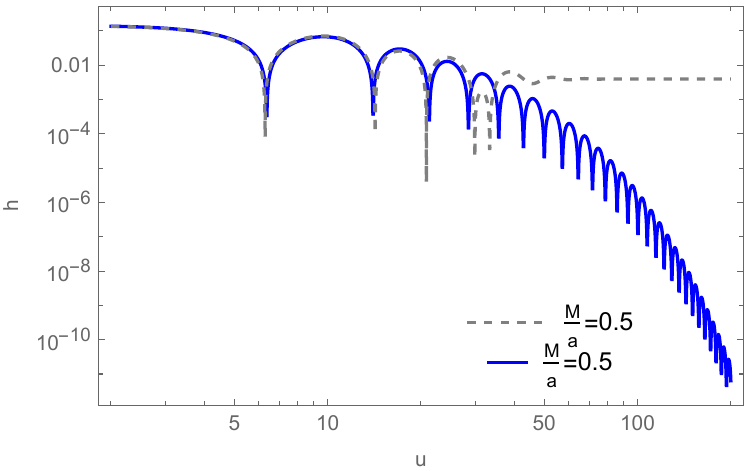}
    \caption{$M/a = 0.5$}
    \label{fig:Memory05}
\end{subfigure}

\vspace{0.3cm}
\begin{subfigure}[b]{0.32\textwidth}
    \centering
    \includegraphics[width=\linewidth]{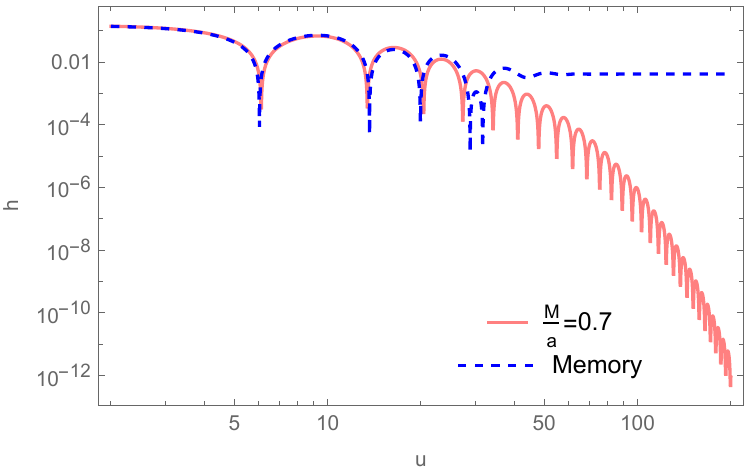}
    \caption{$M/a = 0.7$}
    \label{fig:Memory07}
\end{subfigure}
\hspace{0.02\textwidth}
\begin{subfigure}[b]{0.32\textwidth}
    \centering
    \includegraphics[width=\linewidth]{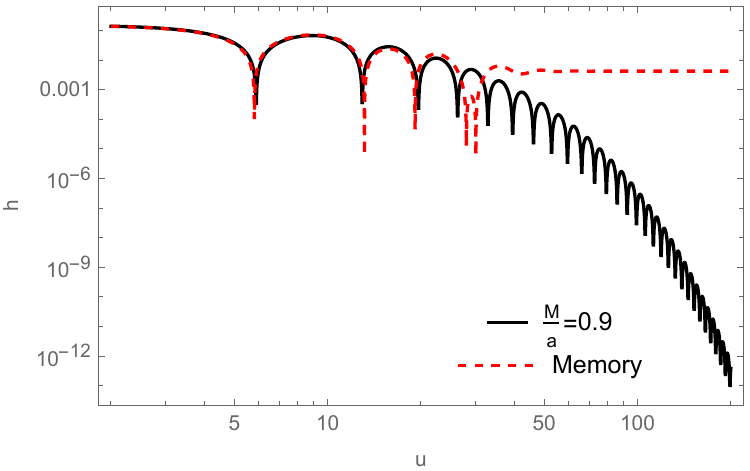}
    \caption{$M/a = 0.9$}
    \label{fig:Memory09}
\end{subfigure}
 \caption{ Log–log plot of the ringdown waveform with and without memory for a black hole surrounded by a dark matter halo, shown for different values of $M/a$. The inclusion of the dark matter halo modifies the late-time behavior of the waveform.}
\label{fig:MemoryAll}
\end{figure*}

\begin{figure*}[htbp]
    \centering
    \begin{subfigure}[b]{0.48\textwidth}
        \centering
  \includegraphics[width=\linewidth]{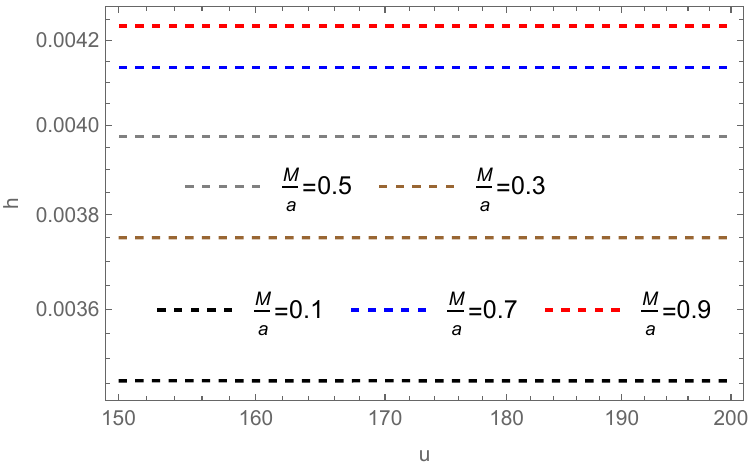}
        \caption{Plot the memory-corrected total waveform, i.e., $ h^{\rm total}$, for different values of $M/a$.}
        \label{fig:Memoryall}
    \end{subfigure}
    \hfill
    \begin{subfigure}[b]{0.45\textwidth}
        \centering
        \includegraphics[width=\linewidth]{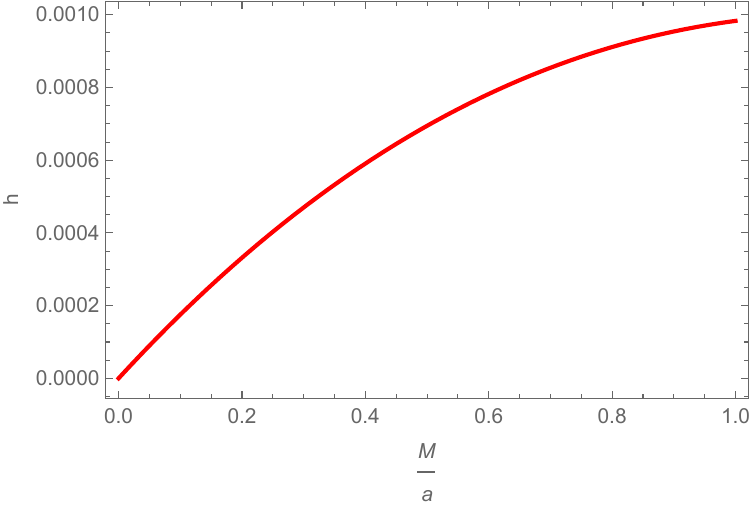}
      \caption{ Relation between the memory-corrected total waveform ($h^{\rm total}$) and $M/a$, given by $h^{\rm total} = 0.002(M/a) - 0.001(M/a)^2 + 0.0001(M/a)^3$.}
        \label{fig:MemoryDM}
    \end{subfigure}
\caption{Dependence of memory corrected total waveform on the dark matter halo profile.}
    \label{fig:MemoryAll_2}
\end{figure*}

For calculating the memory waveform $(h^{\rm memory})$, we use the following relation \cite{Silva:2024ffz} for getting the amplitude of the strain ($h$),
\begin{equation}
    h^{\,(\pm),\,{\rm eff}}_{\ell n} =
    2 \, \left[ \frac{{\rm Re}\; \omega^{\,(\pm)}_{\ell n}}{{\rm Im}\; \omega^{\,(\pm)}_{\ell n}} \right]^{\tfrac{1}{2}} \omega^{\,(\pm)}_{\ell n} \, B^{\,(\pm)}_{\ell n},
\end{equation}

 where $\omega^{(\pm)}_{\ell n}$ denotes the axial-parity and polar-parity quasinormal modes, while $B^{(\pm)}_{\ell n}$ represents the excitation factors of the fundamental quadrupole quasinormal frequencies for these modes. Taking into account the dark matter halo, we compute the quasinormal modes. However, we neglect the halo’s influence on the excitation factors, as our analysis is restricted to the leading order in $M/a$. A more realistic scenario would involve incorporating the effect of dark matter on the excitation factors, which we plan to investigate in the near future. The excitation factor values for $n = 0, 1, 2, 3$ and $\ell = 2$ are taken from \cite{Zhang:2013ksa}. After obtaining the amplitude of the strain $h$, we multiply it by $e^{i \omega_{\ell n} t}$ for $n = 0, 1, 2, 3$ with $\ell = 2$ to construct the strain $h$. Using \ref{memory}, we then compute the memory waveform, and subsequently obtain the memory-corrected total waveform from \ref{tatal_wavefom}. Both the memory-corrected total waveform and the ringdown waveform are shown in \ref{fig:MemoryAll} for various values of $M/a$.  We observe that the memory-corrected total waveform displays a nonlinear dependence on the parameter $M/a$, as illustrated in \ref{fig:Memoryall}. Furthermore, we have obtained the relationship between the memory corrected total waveform $h^{\rm total}$ and $M/a$ using a fitting function as follows,

\begin{equation}
    h^{\rm total}=0.002 (M/a)-0.001 (M/a)^2+0.0001 (M/a)^3,
\end{equation}

\noindent which is plotted in \ref{fig:MemoryDM}. Here, we notice that as $M/a$ increases, the total waveform $h^{\rm total}$ also increases, but the increment rate decreases gradually. What exactly it indicates from the perspective of the dark matter halo is not very clear at this stage and demands further analysis. The memory-corrected total waveform offers a potential pathway to probe the dark matter halo parameter, contingent upon detection sensitivity, an endeavor that could be significantly advanced by next-generation ground-based observatories and, ultimately, by the forthcoming space-based LISA mission.
For constructing memory corrected total waveform templates, we restrict our analysis to the dark matter profile without a spike, as spacetimes with and without a spike exhibit identical asymptotic properties. Consequently, the resulting memory corrected total waveform templates are the same in both cases.




\section{Conclusion} \label{conclusion}

In conclusion, we have presented a detailed study of the GW memory effect for a black hole situated at the center of a galaxy, with particular emphasis on the influence of a surrounding dark matter halo. Our analysis began with the dynamics of test particles in this background, where we solved the geodesic equations in the presence of a GW pulse. This allowed us to explore how the combined effects of the GW perturbation and the halo’s gravitational potential alter both the geodesic deviation equation and the particle trajectories. We found that a passing GW pulse produces a permanent displacement in these quantities, encapsulating the memory effect. In addition, the dark matter halo contributes further corrections, indicating that its gravitational field can meaningfully affect the observable memory signal. From the geodesic analysis, we observed that the memory decreases as $M/a$ increases. Finally, by considering halo profiles with and without a spike, we showed that these cases result in distinct modifications to the particle trajectories. 

To investigate the global structure of the spacetime and the asymptotic behavior of gravitational waves (GWs), we employed the Bondi–Sachs formalism, which is particularly well-suited for characterizing radiative properties at null infinity. Within this framework, we examined the memory effect through the BMS symmetry group, utilizing the associated flux balance laws to derive expressions for the GW memory contribution to the ringdown waveform. For constructing memory corrected total waveform templates, we focused on the dark matter profile without a spike, as spacetimes with and without a spike share identical asymptotic properties. Here, we noticed that as $M/a$ increases, the total waveform increases, but the increment rate decreases gradually. This approach not only provides a covariant and physically transparent description of the memory effect but also establishes a systematic procedure for computing its impact on observables.

Importantly, the incorporation of the BMS flux allows for the encoding of the memory effect into gravitational waveform templates. This paves the way for using GW observations as a novel probe of the astrophysical environment surrounding compact objects. In particular, we highlight the potential of memory-based signatures as a tool to infer properties of dark matter distributions near black holes. As GW detectors improve in sensitivity, the detection of such memory imprints may offer valuable insights into the nature of dark matter and the structure of galactic cores, providing an exciting intersection between GW astronomy and dark matter phenomenology.

\section*{Acknowledgements}
The authors acknowledge helpful discussions with Sumanta Chakraborty, Vitor Cardoso, Keefe Mitman, and Apratim Ganguly.
\bibliography{Blackhol}

\providecommand{\href}[2]{#2}\begingroup\raggedright\begin{thebibliography}{10}

\bibitem{Rubin1970RotationOT}
V.~C. Rubin and W.~K. Ford, ``Rotation of the andromeda nebula from a spectroscopic survey of emission regions,'' {\em The Astrophysical Journal} {\bfseries 159} (1970) 379--403.

\bibitem{cowie1986virial}
L.~L. Cowie, M.~Henriksen, and R.~Mushotzky, ``{Are the Virial Masses of Clusters Smaller Than We Think?},'' \href{http://dx.doi.org/10.1086/165305}{{\em Astrophys. J.} {\bfseries 317} (1987) 593--600}.

\bibitem{Borriello:2000rv}
A.~Borriello and P.~Salucci, ``{The Dark matter distribution in disk galaxies},'' \href{http://dx.doi.org/10.1046/j.1365-8711.2001.04077.x}{{\em Mon. Not. Roy. Astron. Soc.} {\bfseries 323} (2001) 285}, \href{http://arxiv.org/abs/astro-ph/0001082}{{\ttfamily arXiv:astro-ph/0001082}}.

\bibitem{Persic:1995ru}
M.~Persic, P.~Salucci, and F.~Stel, ``{The Universal rotation curve of spiral galaxies: 1. The Dark matter connection},'' \href{http://dx.doi.org/10.1093/mnras/278.1.27}{{\em Mon. Not. Roy. Astron. Soc.} {\bfseries 281} (1996) 27}, \href{http://arxiv.org/abs/astro-ph/9506004}{{\ttfamily arXiv:astro-ph/9506004}}.

\bibitem{Briel:1997hz}
U.~G. Briel and J.~P. Henry, ``{An x-ray temperature map of coma},'' \href{http://arxiv.org/abs/astro-ph/9711237}{{\ttfamily arXiv:astro-ph/9711237}}.

\bibitem{SDSS:2005sxd}
{\bfseries SDSS} Collaboration, J.~K. Adelman-McCarthy {\em et~al.}, ``{The Fourth Data Release of the Sloan Digital Sky Survey},'' \href{http://dx.doi.org/10.1086/497917}{{\em Astrophys. J. Suppl.} {\bfseries 162} (2006) 38--48}, \href{http://arxiv.org/abs/astro-ph/0507711}{{\ttfamily arXiv:astro-ph/0507711}}.

\bibitem{Clowe:2006eq}
D.~Clowe, M.~Bradac, A.~H. Gonzalez, M.~Markevitch, S.~W. Randall, C.~Jones, and D.~Zaritsky, ``{A direct empirical proof of the existence of dark matter},'' \href{http://dx.doi.org/10.1086/508162}{{\em Astrophys. J. Lett.} {\bfseries 648} (2006) L109--L113}, \href{http://arxiv.org/abs/astro-ph/0608407}{{\ttfamily arXiv:astro-ph/0608407}}.

\bibitem{Freese:2008cz}
K.~Freese, ``{Review of Observational Evidence for Dark Matter in the Universe and in upcoming searches for Dark Stars},'' \href{http://dx.doi.org/10.1051/eas/0936016}{{\em EAS Publ. Ser.} {\bfseries 36} (2009) 113--126}, \href{http://arxiv.org/abs/0812.4005}{{\ttfamily arXiv:0812.4005 [astro-ph]}}.

\bibitem{Cardoso:2021wlq}
V.~Cardoso, K.~Destounis, F.~Duque, R.~P. Macedo, and A.~Maselli, ``{Black holes in galaxies: Environmental impact on gravitational-wave generation and propagation},'' \href{http://dx.doi.org/10.1103/PhysRevD.105.L061501}{{\em Phys. Rev. D} {\bfseries 105} no.~6, (2022) L061501}, \href{http://arxiv.org/abs/2109.00005}{{\ttfamily arXiv:2109.00005 [gr-qc]}}.

\bibitem{hernquist1990analytical}
L.~Hernquist, ``An analytical model for spherical galaxies and bulges,'' {\em Astrophysical Journal, Part 1 (ISSN 0004-637X), vol. 356, June 20, 1990, p. 359-364.} {\bfseries 356} (1990) 359--364.

\bibitem{7bb06a79-8225-31c6-88c3-0c4f8a76b072}
A.~Einstein, ``On a stationary system with spherical symmetry consisting of many gravitating masses,'' {\em Annals of Mathematics} {\bfseries 40} no.~4, (1939) 922--936.

\bibitem{Speeney:2024mas}
N.~Speeney, E.~Berti, V.~Cardoso, and A.~Maselli, ``{Black holes surrounded by generic matter distributions: Polar perturbations and energy flux},'' \href{http://dx.doi.org/10.1103/PhysRevD.109.084068}{{\em Phys. Rev. D} {\bfseries 109} no.~8, (2024) 084068}, \href{http://arxiv.org/abs/2401.00932}{{\ttfamily arXiv:2401.00932 [gr-qc]}}.

\bibitem{Chakraborty:2024gcr}
S.~Chakraborty, G.~Comp{\`e}re, and L.~Machet, ``{Tidal Love numbers and quasinormal modes of the Schwarzschild-Hernquist black hole},'' \href{http://dx.doi.org/10.1103/4p2c-rwdh}{{\em Phys. Rev. D} {\bfseries 112} no.~2, (2025) 024015}, \href{http://arxiv.org/abs/2412.14831}{{\ttfamily arXiv:2412.14831 [gr-qc]}}.

\bibitem{LIGOScientific:2016aoc}
{\bfseries LIGO Scientific, Virgo} Collaboration, B.~P. Abbott {\em et~al.}, ``{Observation of Gravitational Waves from a Binary Black Hole Merger},'' \href{http://dx.doi.org/10.1103/PhysRevLett.116.061102}{{\em Phys. Rev. Lett.} {\bfseries 116} no.~6, (2016) 061102}, \href{http://arxiv.org/abs/1602.03837}{{\ttfamily arXiv:1602.03837 [gr-qc]}}.

\bibitem{LIGOScientific:2017vwq}
{\bfseries LIGO Scientific, Virgo} Collaboration, B.~P. Abbott {\em et~al.}, ``{GW170817: Observation of Gravitational Waves from a Binary Neutron Star Inspiral},'' \href{http://dx.doi.org/10.1103/PhysRevLett.119.161101}{{\em Phys. Rev. Lett.} {\bfseries 119} no.~16, (2017) 161101}, \href{http://arxiv.org/abs/1710.05832}{{\ttfamily arXiv:1710.05832 [gr-qc]}}.

\bibitem{KAGRA:2025oiz}
{\bfseries KAGRA, Virgo, LIGO Scientific} Collaboration, A.~G. Abac {\em et~al.}, ``{GW250114: Testing Hawking{\textquoteright}s Area Law and the Kerr Nature of Black Holes},'' \href{http://dx.doi.org/10.1103/kw5g-d732}{{\em Phys. Rev. Lett.} {\bfseries 135} no.~11, (2025) 111403}, \href{http://arxiv.org/abs/2509.08054}{{\ttfamily arXiv:2509.08054 [gr-qc]}}.

\bibitem{EventHorizonTelescope:2019dse}
{\bfseries Event Horizon Telescope} Collaboration, K.~Akiyama {\em et~al.}, ``{First M87 Event Horizon Telescope Results. I. The Shadow of the Supermassive Black Hole},'' \href{http://dx.doi.org/10.3847/2041-8213/ab0ec7}{{\em Astrophys. J. Lett.} {\bfseries 875} (2019) L1}, \href{http://arxiv.org/abs/1906.11238}{{\ttfamily arXiv:1906.11238 [astro-ph.GA]}}.

\bibitem{EventHorizonTelescope:2019jan}
{\bfseries Event Horizon Telescope} Collaboration, K.~Akiyama {\em et~al.}, ``{First M87 Event Horizon Telescope Results. III. Data Processing and Calibration},'' \href{http://dx.doi.org/10.3847/2041-8213/ab0c57}{{\em Astrophys. J. Lett.} {\bfseries 875} no.~1, (2019) L3}, \href{http://arxiv.org/abs/1906.11240}{{\ttfamily arXiv:1906.11240 [astro-ph.GA]}}.

\bibitem{EventHorizonTelescope:2019pgp}
{\bfseries Event Horizon Telescope} Collaboration, K.~Akiyama {\em et~al.}, ``{First M87 Event Horizon Telescope Results. V. Physical Origin of the Asymmetric Ring},'' \href{http://dx.doi.org/10.3847/2041-8213/ab0f43}{{\em Astrophys. J. Lett.} {\bfseries 875} no.~1, (2019) L5}, \href{http://arxiv.org/abs/1906.11242}{{\ttfamily arXiv:1906.11242 [astro-ph.GA]}}.

\bibitem{EventHorizonTelescope:2019ths}
{\bfseries Event Horizon Telescope} Collaboration, K.~Akiyama {\em et~al.}, ``{First M87 Event Horizon Telescope Results. IV. Imaging the Central Supermassive Black Hole},'' \href{http://dx.doi.org/10.3847/2041-8213/ab0e85}{{\em Astrophys. J. Lett.} {\bfseries 875} no.~1, (2019) L4}, \href{http://arxiv.org/abs/1906.11241}{{\ttfamily arXiv:1906.11241 [astro-ph.GA]}}.

\bibitem{EventHorizonTelescope:2019uob}
{\bfseries Event Horizon Telescope} Collaboration, K.~Akiyama {\em et~al.}, ``{First M87 Event Horizon Telescope Results. II. Array and Instrumentation},'' \href{http://dx.doi.org/10.3847/2041-8213/ab0c96}{{\em Astrophys. J. Lett.} {\bfseries 875} no.~1, (2019) L2}, \href{http://arxiv.org/abs/1906.11239}{{\ttfamily arXiv:1906.11239 [astro-ph.IM]}}.

\bibitem{Boersma:2020gxx}
O.~M. Boersma, D.~A. Nichols, and P.~Schmidt, ``{Forecasts for detecting the gravitational-wave memory effect with Advanced LIGO and Virgo},'' \href{http://dx.doi.org/10.1103/PhysRevD.101.083026}{{\em Phys. Rev. D} {\bfseries 101} no.~8, (2020) 083026}, \href{http://arxiv.org/abs/2002.01821}{{\ttfamily arXiv:2002.01821 [astro-ph.HE]}}.

\bibitem{Braginsky:1985vlg}
V.~B. Braginsky and L.~P. Grishchuk, ``{Kinematic Resonance and Memory Effect in Free Mass Gravitational Antennas},'' {\em Sov. Phys. JETP} {\bfseries 62} (1985) 427--430.

\bibitem{Favata:2010zu}
M.~Favata, ``{The gravitational-wave memory effect},'' \href{http://dx.doi.org/10.1088/0264-9381/27/8/084036}{{\em Class. Quant. Grav.} {\bfseries 27} (2010) 084036}, \href{http://arxiv.org/abs/1003.3486}{{\ttfamily arXiv:1003.3486 [gr-qc]}}.

\bibitem{Hubner:2021amk}
M.~H\"ubner, P.~Lasky, and E.~Thrane, ``{Memory remains undetected: Updates from the second LIGO/Virgo gravitational-wave transient catalog},'' \href{http://dx.doi.org/10.1103/PhysRevD.104.023004}{{\em Phys. Rev. D} {\bfseries 104} no.~2, (2021) 023004}, \href{http://arxiv.org/abs/2105.02879}{{\ttfamily arXiv:2105.02879 [gr-qc]}}.

\bibitem{Lasky:2016knh}
P.~D. Lasky, E.~Thrane, Y.~Levin, J.~Blackman, and Y.~Chen, ``{Detecting gravitational-wave memory with LIGO: implications of GW150914},'' \href{http://dx.doi.org/10.1103/PhysRevLett.117.061102}{{\em Phys. Rev. Lett.} {\bfseries 117} no.~6, (2016) 061102}, \href{http://arxiv.org/abs/1605.01415}{{\ttfamily arXiv:1605.01415 [astro-ph.HE]}}.

\bibitem{Zeldovich:1974gvh}
Y.~B. Zel'dovich and A.~G. Polnarev, ``{Radiation of gravitational waves by a cluster of superdense stars},'' {\em Sov. Astron.} {\bfseries 18} (1974) 17.

\bibitem{Kovacs:1978eu}
S.~J. Kovacs and K.~S. Thorne, ``{The Generation of Gravitational Waves. 4. Bremsstrahlung},'' \href{http://dx.doi.org/10.1086/156350}{{\em Astrophys. J.} {\bfseries 224} (1978) 62--85}.

\bibitem{Christodoulou:1991cr}
D.~Christodoulou, ``{Nonlinear nature of gravitation and gravitational wave experiments},'' \href{http://dx.doi.org/10.1103/PhysRevLett.67.1486}{{\em Phys. Rev. Lett.} {\bfseries 67} (1991) 1486--1489}.

\bibitem{Bieri:2013hqa}
L.~Bieri and D.~Garfinkle, ``{An electromagnetic analogue of gravitational wave memory},'' \href{http://dx.doi.org/10.1088/0264-9381/30/19/195009}{{\em Class. Quant. Grav.} {\bfseries 30} (2013) 195009}, \href{http://arxiv.org/abs/1307.5098}{{\ttfamily arXiv:1307.5098 [gr-qc]}}.

\bibitem{Winicour:2014ska}
J.~Winicour, ``{Global aspects of radiation memory},'' \href{http://dx.doi.org/10.1088/0264-9381/31/20/205003}{{\em Class. Quant. Grav.} {\bfseries 31} (2014) 205003}, \href{http://arxiv.org/abs/1407.0259}{{\ttfamily arXiv:1407.0259 [gr-qc]}}.

\bibitem{Pate:2017vwa}
M.~Pate, A.-M. Raclariu, and A.~Strominger, ``{Color Memory: A Yang-Mills Analog of Gravitational Wave Memory},'' \href{http://dx.doi.org/10.1103/PhysRevLett.119.261602}{{\em Phys. Rev. Lett.} {\bfseries 119} no.~26, (2017) 261602}, \href{http://arxiv.org/abs/1707.08016}{{\ttfamily arXiv:1707.08016 [hep-th]}}.

\bibitem{Jokela:2019apz}
N.~Jokela, K.~Kajantie, and M.~Sarkkinen, ``{Memory effect in Yang-Mills theory},'' \href{http://dx.doi.org/10.1103/PhysRevD.99.116003}{{\em Phys. Rev. D} {\bfseries 99} no.~11, (2019) 116003}, \href{http://arxiv.org/abs/1903.10231}{{\ttfamily arXiv:1903.10231 [hep-th]}}.

\bibitem{Hollands:2016oma}
S.~Hollands, A.~Ishibashi, and R.~M. Wald, ``{BMS Supertranslations and Memory in Four and Higher Dimensions},'' \href{http://dx.doi.org/10.1088/1361-6382/aa777a}{{\em Class. Quant. Grav.} {\bfseries 34} no.~15, (2017) 155005}, \href{http://arxiv.org/abs/1612.03290}{{\ttfamily arXiv:1612.03290 [gr-qc]}}.

\bibitem{Satishchandran:2017pek}
G.~Satishchandran and R.~M. Wald, ``{Memory effect for particle scattering in odd spacetime dimensions},'' \href{http://dx.doi.org/10.1103/PhysRevD.97.024036}{{\em Phys. Rev. D} {\bfseries 97} no.~2, (2018) 024036}, \href{http://arxiv.org/abs/1712.00873}{{\ttfamily arXiv:1712.00873 [gr-qc]}}.

\bibitem{Ferko:2021bym}
C.~Ferko, G.~Satishchandran, and S.~Sethi, ``{Gravitational memory and compact extra dimensions},'' \href{http://dx.doi.org/10.1103/PhysRevD.105.024072}{{\em Phys. Rev. D} {\bfseries 105} no.~2, (2022) 024072}, \href{http://arxiv.org/abs/2109.11599}{{\ttfamily arXiv:2109.11599 [gr-qc]}}.

\bibitem{Du:2016hww}
S.~M. Du and A.~Nishizawa, ``{Gravitational Wave Memory: A New Approach to Study Modified Gravity},'' \href{http://dx.doi.org/10.1103/PhysRevD.94.104063}{{\em Phys. Rev. D} {\bfseries 94} no.~10, (2016) 104063}, \href{http://arxiv.org/abs/1609.09825}{{\ttfamily arXiv:1609.09825 [gr-qc]}}.

\bibitem{Hou:2020tnd}
S.~Hou and Z.-H. Zhu, ``{Gravitational memory effects and Bondi-Metzner-Sachs symmetries in scalar-tensor theories},'' \href{http://dx.doi.org/10.1007/JHEP01(2021)083}{{\em JHEP} {\bfseries 01} (2021) 083}, \href{http://arxiv.org/abs/2005.01310}{{\ttfamily arXiv:2005.01310 [gr-qc]}}.

\bibitem{Hou:2020xme}
S.~Hou, ``{Gravitational memory effects in Brans-Dicke theory},'' \href{http://dx.doi.org/10.1002/asna.202113887}{{\em Astron. Nachr.} {\bfseries 342} no.~1-2, (2021) 96--102}, \href{http://arxiv.org/abs/2011.02087}{{\ttfamily arXiv:2011.02087 [gr-qc]}}.

\bibitem{Tahura:2020vsa}
S.~Tahura, D.~A. Nichols, A.~Saffer, L.~C. Stein, and K.~Yagi, ``{Brans-Dicke theory in Bondi-Sachs form: Asymptotically flat solutions, asymptotic symmetries and gravitational-wave memory effects},'' \href{http://dx.doi.org/10.1103/PhysRevD.103.104026}{{\em Phys. Rev. D} {\bfseries 103} no.~10, (2021) 104026}, \href{http://arxiv.org/abs/2007.13799}{{\ttfamily arXiv:2007.13799 [gr-qc]}}.

\bibitem{Hou:2021oxe}
S.~Hou, T.~Zhu, and Z.-H. Zhu, ``{Asymptotic analysis of Chern-Simons modified gravity and its memory effects},'' \href{http://dx.doi.org/10.1103/PhysRevD.105.024025}{{\em Phys. Rev. D} {\bfseries 105} no.~2, (2022) 024025}, \href{http://arxiv.org/abs/2109.04238}{{\ttfamily arXiv:2109.04238 [gr-qc]}}.

\bibitem{Bhattacharya:2022}
I.~Chakraborty, S.~Bhattacharya, and S.~Chakraborty, ``Gravitational wave memory in wormhole spacetimes,'' \href{http://dx.doi.org/10.1103/PhysRevD.106.104057}{{\em Phys. Rev. D} {\bfseries 106} (Nov, 2022) 104057}. \url{https://link.aps.org/doi/10.1103/PhysRevD.106.104057}.

\bibitem{bhattacharya:2023}
S.~Bhattacharya and S.~Ghosh, ``Gravitational wave memory for a class of static and spherically symmetric spacetimes,'' 2023.
\newblock \url{https://arxiv.org/abs/2309.04130}.

\bibitem{bhattacharya:2025}
S.~Bhattacharya and S.~Ghosh, ``Displacement memory and b-memory in generalised ellis-bronnikov wormholes,'' 2025.
\newblock \url{https://arxiv.org/abs/2502.03007}.

\bibitem{Bhattacharya_2024}
S.~Bhattacharya, D.~Bose, I.~Chakraborty, A.~Hait, and S.~Mohanty, ``{Gravitational memory signal from neutrino self-interactions in supernova},'' \href{http://dx.doi.org/10.1103/PhysRevD.110.L061501}{{\em Phys. Rev. D} {\bfseries 110} no.~6, (2024) L061501}, \href{http://arxiv.org/abs/2311.03315}{{\ttfamily arXiv:2311.03315 [gr-qc]}}.

\bibitem{Zhang:2017}
P.-M. Zhang, C.~Duval, G.~W. Gibbons, and P.~A. Horvathy, ``{The Memory Effect for Plane Gravitational Waves},'' \href{http://dx.doi.org/10.1016/j.physletb.2017.07.050}{{\em Phys. Lett. B} {\bfseries 772} (2017) 743--746}.

\bibitem{Zhang:2017soft}
P.-M. Zhang, C.~Duval, G.~W. Gibbons, and P.~A. Horvathy, ``{Soft gravitons and the memory effect for plane gravitational waves},'' \href{http://dx.doi.org/10.1103/PhysRevD.96.064013}{{\em Phys. Rev. D} {\bfseries 96} no.~6, (2017) 064013}.

\bibitem{Horvathy_2017}
P.~M. Zhang, C.~Duval, G.~W. Gibbons, and P.~A. Horvathy, ``{Soft gravitons and the memory effect for plane gravitational waves},'' \href{http://dx.doi.org/10.1103/PhysRevD.96.064013}{{\em Phys. Rev. D} {\bfseries 96} no.~6, (2017) 064013}, \href{http://arxiv.org/abs/1705.01378}{{\ttfamily arXiv:1705.01378 [gr-qc]}}.

\bibitem{Donnay:2018ckb}
L.~Donnay, G.~Giribet, H.~A. Gonz\'alez, and A.~Puhm, ``{Black hole memory effect},'' \href{http://dx.doi.org/10.1103/PhysRevD.98.124016}{{\em Phys. Rev. D} {\bfseries 98} no.~12, (2018) 124016}, \href{http://arxiv.org/abs/1809.07266}{{\ttfamily arXiv:1809.07266 [hep-th]}}.

\bibitem{Bhattacharjee:2019jaf}
S.~Bhattacharjee, S.~Kumar, and A.~Bhattacharyya, ``{Memory Effect and BMS-like Symmetries for Impulsive Gravitational Waves},'' \href{http://dx.doi.org/10.1103/PhysRevD.100.084010}{{\em Phys. Rev. D} {\bfseries 100} no.~8, (2019) 084010}, \href{http://arxiv.org/abs/1905.12905}{{\ttfamily arXiv:1905.12905 [hep-th]}}.

\bibitem{Rahman:2019bmk}
A.~A. Rahman and R.~M. Wald, ``{Black Hole Memory},'' \href{http://dx.doi.org/10.1103/PhysRevD.101.124010}{{\em Phys. Rev. D} {\bfseries 101} no.~12, (2020) 124010}, \href{http://arxiv.org/abs/1912.12806}{{\ttfamily arXiv:1912.12806 [gr-qc]}}.

\bibitem{Alnasheet:2025mtr}
Q.~Alnasheet, V.~Cardoso, F.~Duque, and R.~Panosso~Macedo, ``{Gravitational-wave tails and memory effect for mergers in astrophysical environments},'' \href{http://arxiv.org/abs/2508.20238}{{\ttfamily arXiv:2508.20238 [gr-qc]}}.

\bibitem{Zhang:2017geq}
P.~M. Zhang, C.~Duval, G.~W. Gibbons, and P.~A. Horvathy, ``{Soft gravitons and the memory effect for plane gravitational waves},'' \href{http://dx.doi.org/10.1103/PhysRevD.96.064013}{{\em Phys. Rev. D} {\bfseries 96} no.~6, (2017) 064013}, \href{http://arxiv.org/abs/1705.01378}{{\ttfamily arXiv:1705.01378 [gr-qc]}}.

\bibitem{Madler:2016xju}
T.~M\"adler and J.~Winicour, ``{Bondi-Sachs Formalism},'' \href{http://dx.doi.org/10.4249/scholarpedia.33528}{{\em Scholarpedia} {\bfseries 11} (2016) 33528}, \href{http://arxiv.org/abs/1609.01731}{{\ttfamily arXiv:1609.01731 [gr-qc]}}.

\bibitem{Strominger:2017zoo}
A.~Strominger, {\em {Lectures on the Infrared Structure of Gravity and Gauge Theory}}.
\newblock 3, 2017.
\newblock \href{http://arxiv.org/abs/1703.05448}{{\ttfamily arXiv:1703.05448 [hep-th]}}.

\bibitem{Ashtekar_2020}
A.~Ashtekar, T.~De~Lorenzo, and N.~Khera, ``{Compact binary coalescences: Constraints on waveforms},'' \href{http://dx.doi.org/10.1007/s10714-020-02764-1}{{\em Gen. Rel. Grav.} {\bfseries 52} no.~11, (2020) 107}, \href{http://arxiv.org/abs/1906.00913}{{\ttfamily arXiv:1906.00913 [gr-qc]}}.

\bibitem{Mitman_2020}
K.~Mitman, J.~Moxon, M.~A. Scheel, S.~A. Teukolsky, M.~Boyle, N.~Deppe, L.~E. Kidder, and W.~Throwe, ``{Computation of displacement and spin gravitational memory in numerical relativity},'' \href{http://dx.doi.org/10.1103/PhysRevD.102.104007}{{\em Phys. Rev. D} {\bfseries 102} no.~10, (2020) 104007}, \href{http://arxiv.org/abs/2007.11562}{{\ttfamily arXiv:2007.11562 [gr-qc]}}.

\bibitem{Mitman_2021}
K.~Mitman {\em et~al.}, ``{Adding gravitational memory to waveform catalogs using BMS balance laws},'' \href{http://dx.doi.org/10.1103/PhysRevD.103.024031}{{\em Phys. Rev. D} {\bfseries 103} no.~2, (2021) 024031}, \href{http://arxiv.org/abs/2011.01309}{{\ttfamily arXiv:2011.01309 [gr-qc]}}.

\bibitem{deppe2025}
N.~Deppe, L.~Heisenberg, L.~E. Kidder, D.~Maibach, S.~Ma, J.~Moxon, K.~C. Nelli, W.~Throwe, and N.~L. Vu, ``Signatures of quantum gravity in gravitational wave memory,'' 2025.
\newblock \url{https://arxiv.org/abs/2502.20584}.

\bibitem{Moxon_2020}
J.~Moxon, M.~A. Scheel, and S.~A. Teukolsky, ``{Improved Cauchy-characteristic evolution system for high-precision numerical relativity waveforms},'' \href{http://dx.doi.org/10.1103/PhysRevD.102.044052}{{\em Phys. Rev. D} {\bfseries 102} no.~4, (2020) 044052}, \href{http://arxiv.org/abs/2007.01339}{{\ttfamily arXiv:2007.01339 [gr-qc]}}.

\bibitem{Silva:2024ffz}
H.~O. Silva, G.~Tambalo, K.~Glampedakis, K.~Yagi, and J.~Steinhoff, ``{Quasinormal modes and their excitation beyond general relativity},'' \href{http://dx.doi.org/10.1103/PhysRevD.110.024042}{{\em Phys. Rev. D} {\bfseries 110} no.~2, (2024) 024042}, \href{http://arxiv.org/abs/2404.11110}{{\ttfamily arXiv:2404.11110 [gr-qc]}}.

\bibitem{Zhang:2013ksa}
Z.~Zhang, E.~Berti, and V.~Cardoso, ``{Quasinormal ringing of Kerr black holes. II. Excitation by particles falling radially with arbitrary energy},'' \href{http://dx.doi.org/10.1103/PhysRevD.88.044018}{{\em Phys. Rev. D} {\bfseries 88} (2013) 044018}, \href{http://arxiv.org/abs/1305.4306}{{\ttfamily arXiv:1305.4306 [gr-qc]}}.

\end{thebibliography}\endgroup

\bibliographystyle{./utphys1}


\end{document}